# Enhancement in Electro-Optic Properties of Dynamic Scattering Systems through Addition of Dichroic Dyes

Mansoureh Shasti[1], and James. T. Gleeson[1]

*Dept of Physics, Kent State University, Kent, OH 44242*

Paul Luchette[2], Tamas Kosa[2], Antonio F. Munoz[2], and Bahman Taheri[2]

*AlphaMicron, Inc, 1950 State Route 59, Kent, OH*

**Abstract:** Electro-optic properties of dynamic scattering in homeotropically aligned pure and dichroic dye doped nematic liquid crystal samples are examined. The optical properties of the two systems are quantified using transmission properties of scattered and unscattered as a function of amplitude and frequency of an applied voltage. Auto-correlation of the scattered signal at different applied voltages is used to compare the decay times in the two systems. Lastly, histogram of the scattered signal reveals a wavevector dependent large light scattering event. The dye doped system shows a significant enhancement of light blocking property in both normal and off-axis light propagation. The characteristics of the system are compared to other scattering technologies. The results suggest that dye doped system can overcome shortcomings in scattering based devices used for near eye applications.

## 1. Introduction

Dynamic scattering mode (DSM), invented in 1968, is one of the oldest liquid crystal based technologies for light management [1-4]. Like its modern contemporaries, namely PDLC, NCAP, and PSCT, the main mechanism for light management is through field induced light scattering. Unlike the others, DSM relies on hydrodynamics of liquid crystals and is based on turbulent fluid motion rising from field-induced flow instabilities [5,6]. The effect of turbulence on the nematic director (direction of optical anisotropy) creates large, random gradients in refractive index which strongly scatter light [7-10]. This system has the distinct advantage over the other scattering technologies in that it can offer a high level of optical clarity (low haze) in the off-state. Its performance is reliant on presence of conductivity in the mixture and hence the composition of the liquid crystalline material used is critical to performance.

One key shortcoming of scattering based liquid crystal devices is that light scattering is primarily in the forward direction. This reduces the overall light blockage characteristics and can lead to parasitic effects in bright sunlight. To mitigate this, absorbing dyes are sometimes added to absorb as well as scatter the ambient light. Much like the other scattering systems, addition of dichroic dyes to dynamic scattering also improves the optical performance.

Furthermore, the low haze/high clarity of DSM in the off state reduces the effect of the dye in the clear state is reduced. However, due to the strong material composition dependence of DSM scattering, the exact effects of adding dyes are not fully predictable and have not been explored. Here, we examine a dye doped DSM composed of a nematic liquid crystal host combined with a dichroic dye guest [11,12]. The dichroic dyes used exhibit positive dichroism wherein the effective absorption dipole of the dye is parallel to the nematic director [13]. The electro-optic performance of this systems is compared to an undoped DSM.

The results are then used to provide a basis for comparison with other scattering technologies for potential use for near eye applications.

1. **Experimental:**

The guest-host DSM (GH-DSM) is comprised of a negative dielectric constant nematic liquid crystal host EMD 2159 from Merck KGaA, and black dichroic dye SLO-6609 from AlphaMicron Inc. and BenzylDimthylhexadecyl Ammonium Chloride (BDAC) additive to increase electrical conductivity [14]. The formulated mixture has a positive conductivity anisotropy and negative dielectric anisotropy as required to exhibit the dynamic scattering mode. This material was sandwiched between two planar, ITO coated transparent substrates that serve as electrodes by 10$\mu$m sphere spacers. The surfaces are pre-treated to produce a uniform homeotropic alignment to give maximum transmissivity in the off (zero voltage) state.

For all measurements, the electric field was produced by a function generator producing a square AC wave (having variable amplitude and frequency). Except as noted, the frequency was set at 100Hz, well below the cut-off frequency ($f_c$ – described below) to ensure the system remained in the conductive regime. The applied potential difference and frequency was monitored using a digital scope in parallel with the device.

The instability diagrams were created by viewing each sample device using polarizing light microscopy to observe the onset of convective rolls and subsequent higher-order instabilities. The voltage was then increased at a given fixed frequency to establish the phase diagram.

For unscattered light transmittance measurements, an incandescent light source, focused on the sample device. The transmitted light beam was then focused by a secondary lens into a fiber and directed to the Ocean Optics spectrometer for processing. The transmission at 550 nm was selected as the benchmark for comparison between sample devices.

Measurement of light transmission in the scattering state is more complex. The results are dependent of the geometry of the setup. This is further complicated if the system is to be used for application where human judgment is required. Therefore, to quantify the observations in a reproducible way, two methods were employed. The first is a traditional low angle deviation measurement of scattered light. The experimental setup for this measurement is shown in Figure 1. The light source used is a 15mW, linearly polarized He-Ne laser with 632.8 nm wavelength. An adjustable iris is used to define the beam waist at the sample. The scattered light is captured by a photomultiplier tube which is oriented at small angle (0.8º) to the direct beam to sample the scattered light. The output of the detector is pre-amplified, filtered and interfaced to the computer via an analog-digital converter. Labview was used to control the amplitude and frequency of the applied voltage, and to collect and analyze the scattered light intensity data. This program performs the autocorrelation, Fast Fourier Transform (FFT) and histogram analysis.

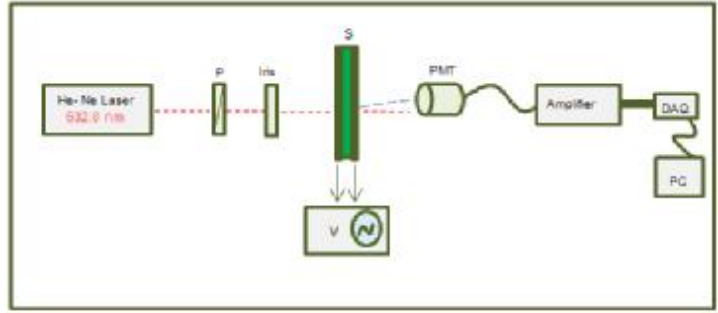

Fig. 1: Experimental setup for measurement of scattered light. He-Ne lase, Polarizer, Iris, sample with the voltage supplier, PMT detector, pre-amplifier, DAQ system to convert the analog data to digital one and the computer.

For overall scattering property, image degradation methods using a negative, back illuminated 50.8mmx50.8mm USAF 1951 test target and an SLR camera with 50mm lens are used. The test target has 26 line sets each comprising of five horizontal and five vertical lines that is labeled with the frequency of the lines in cycles/mm. (Figure 2) In order to make uniform comparisons, all photos were captured using the same shutter and aperture settings.

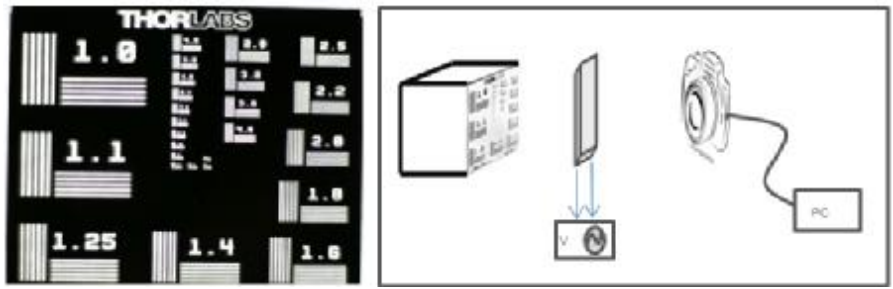

Fig. 2: Test target (Left picture) and the target set up (Right picture).

## 2. Experimental Results and discussions:

Figure 3 shows representative photographs of the GH-DSM and DSM samples in the two extreme conditions of fully OFF and fully ON states. With no applied voltage, the devices are topically clear (high optical clarity). It should be noted that GH-DSM samples exhibit a clear state absorption due to the presence of the added dichroic dye. At 20 V both samples exhibit turbulent dynamic scattering and are opaque. Samples without dye have a white appearance characteristic of strong scattering of ambient light. The GH-DSM samples is cloudy and grey/black color, indicating that the ambient light has absorption as well as scattering.

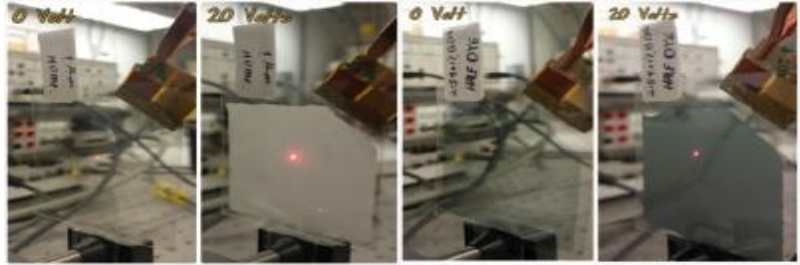

Fig. 3: The cells at zero and 20 Volts (Off and On states) Right: without dye; Left, containing dye.

The first step is to measure the threshold voltage for convective rolls as a function of applied frequency. The instability diagram of each cell has two regimes (typically known as dielectric and conductive) which are separated by the cut-off frequency $f_c$.

The stability diagram establishes the operating parameters for these devices. This diagram, shown in Figure 4 and 5 reveal the phase diagrams and threshold voltage for rolling pattern versus frequency, respectively.

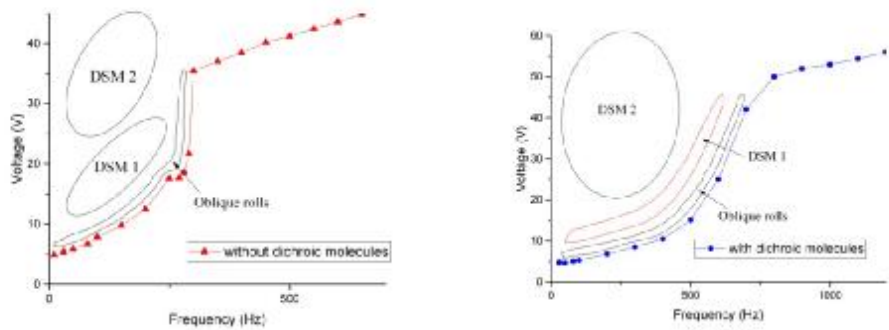

Fig. 4: Phase diagrams for cells without dichroic and with dichroic molecules.

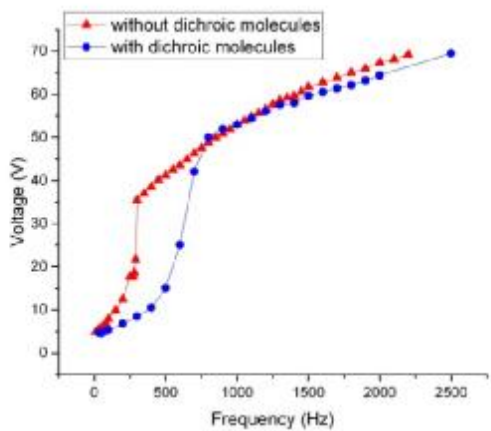

Fig. 5: Threshold voltage for rolling pattern versus frequency for cells without dichroic and with dichroic molecules.

We measured the threshold voltage to see the onset of electroconvection versus frequency. As the voltage is increased above threshold, the convective rolls become more chaotic, ultimately transitioning into the dynamic scattering mode (DSM). The cut-off frequency is 300 Hz and 800 Hz for without and with dichroic molecules cell, respectively. The mixtures with dichroic molecules exhibit a higher conductivity which in turn leads to a larger cutoff frequency. For both cells (with and without dichroic molecules) the cut-off frequency is high enough to ensure that the DSM for the eye wear application is in the conductive regime. The strongest scattering state is dynamic scattering mode 2 (DSM-2). This regime is always attained with operating frequency 100Hz, and applied voltage above 20V.

The direct light transmission of these devices can be adjusted from 80-90% in the "off" state to essentially zero in the DSM-2 state. See Figure 6. The device without dye remains mostly transparent until the onset of electroconvection at around 8 V, while the device containing dye shows decreased transmission as soon as the Freedericksz transition occurs at 2-3V as the guest-host effect occurs. The transmission of this device continues to decrease as the electroconvective instabilities are triggered when the voltage is increased. At 12 V, this device is almost opaque. These data demonstrate how the light transmission can be continually adjusted by applying voltage to the cells for eyewear applications.

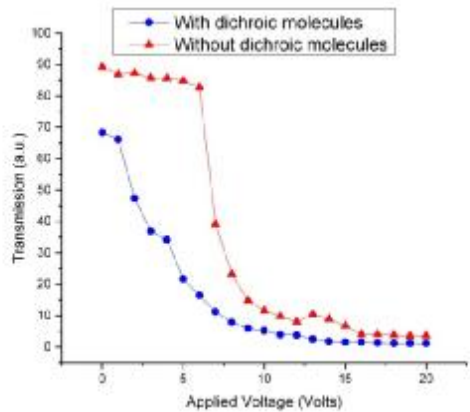

Fig. 6: Transmission versus applied voltage for cells with and without dichroic molecules.

The transmission of light after passing through the cell decreases by applying voltage for both sample cells. The comparison between the cell with dichroic molecules (Red circle) and the cell without dichroic molecules (Black circle) shows that the cell with dichroic molecules absorb more due to the polarizer being parallel to molecules, so the transmission is even less than the cell without dichroic molecules. So we can decrease the transmission 90% by increasing applied voltage which is perfect for lots of applications especially our main goal for the eye wear application.

The scattering intensity is a measure of how much light is deviated from its original direction. This is shown in Figure 7. We also measured the averaged scattered light versus voltage for cells with and without dichroic molecules. They both show low scattering at zero voltage (where the devices are transparent), and low scattering intensity at high voltage (where the devices are almost opaque). The device containing dichroic dye shows a greatly reduced level of scattering at intermediate voltages in DSM2, they have almost same average intensity values.

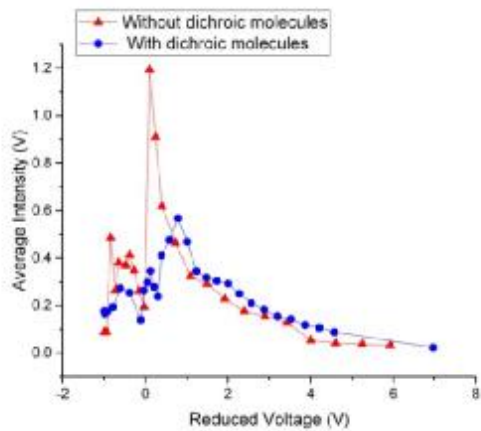

Fig. 7: Scattered light Intensity versus reduced applied voltage for cells with and without dichroic molecules.

Because vision is sensitive to not only the overall amount of light, but also time-dependent changes in light, it is critical that eyewear devices not distract the wearer. Since DSM devices rely on time-dependent scattering of light, they have the potential to become distracting if the character of scattered light is either too intense (specular scattering) or too persistent. For that reason we turned to statistical analyses of scattered light.

The autocorrelation is the correlation of the signal is essentially a measure of to what extent the past values of a fluctuating quantity can predict the present. For our purposes, if the autocorrelation decays quickly enough with time, then temporal fluctuations in scattered light will not last long enough to distract the wearer's vision.

The auto-correlation as $g(t)$ could be defined for a signal:

$$g(t) = \frac{1}{T}\int_0^T I(t)I(t-t)dt, \qquad (1)$$

Where $I(t)$ is the normalized light intensity fluctuation from the mean [15].

Figure 8 shows a few examples of the autocorrelation function at differing applied voltages. Note that at very short values of the time $t$ the autocorrelation is finite, but then fluctuates around zero for longer times. Moreover, at higher applied voltages, the autocorrelation decays to zero in shorter times. The decay time is the time interval needed for this to happen.

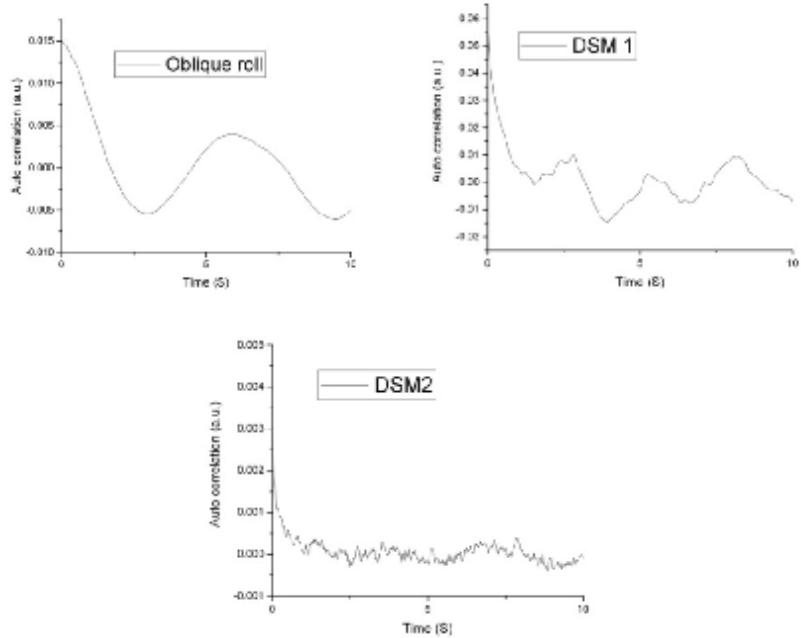

Fig. 8: Auto correlation amplitude versus time for a cell with dichroic molecules in oblique roll, DSM1 and DSM2 modes. The frequency is 100 Hz and it is filtered at 30 Hz. Note that in the oblique mode the autocorrelation reflects the oscillatory nature of the electroconvection pattern.

In order to make meaningful distinction between different devices, the autocorrelation decay time is presented as a function of reduced voltage $(V^2/V_c^2)-1$, where $V_c$ is the threshold voltage for convective rolls.) The decay time is estimated by via linear fit of the autocorrelation function on a semi logarithmic plot (Figure 9). The correlation time decreases by increasing voltage so we have faster decay for higher voltage. The correlation time for very scattered signal (DSM) is sufficiently less than typical visual response which enables a scattering device with no detectable flicker.

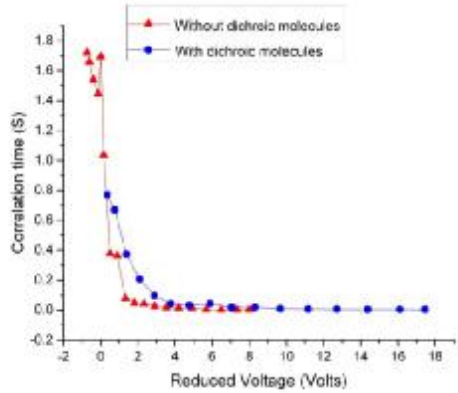

Fig. 9: Correlation time versus reduced voltage for cells without dichroic and with dichroic molecules.

With the autocorrelation analysis we are able to characterize the dynamic properties of the scattering with regards to the potential for light fluctuations that have lifetime great enough to be perceived. In addition, we also must control for anomalously bright intensity fluctuations, even if they are short-lived. To that end, we analyze the probability of bright fluctuations, with a view to reducing the outlier frequency. Figure 10 shows a semi-logarithmic histogram plot of scattered light intensity for two embodiments in DSM2 regime: with and without dichroic dye. While for both embodiments, the frequency of lower than average fluctuations is similar, we see a marked decrease in the frequency of fluctuations much brighter than average. Specifically, brief, bright fluctuations occur about $3x$ less often in the sample containing dichroic dye. Thus, this preparation better addresses the needs for eyewear applications.

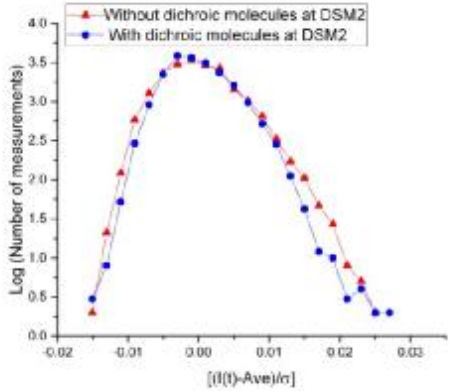

Fig. 10: Semi-logarithmic histogram plot of scattered light intensity for two embodiments in DSM2 regime: with and without dichroic dye.

## 3. Application

These devices are extremely effective at reducing visual acuity which still exhibiting a clear state transmission of >60%. This is demonstrated in Figures 11 and 12 Photographs of a standard resolution target (taken under identical conditions) show that the resolution is degraded significantly (less than 1.4 L/mm) at 8 V operating conditions in the cell containing dye. At 16 V, the target is completely obscured. Furthermore, as shown in Figure 11 the cell with dichroic molecules at different voltages (0, 8 and 12 volts) and achieve the higher opacity at a much lower voltage. The same process for the cell without dichroic molecules at 0, 8, 12, 16 and 20 volts (Figure 12) but the transmission light will be completely blocked at 20 volts. Furthermore, the cell without dye exhibits the parasitic off angle scattering in brightly illuminated lighting conditions.

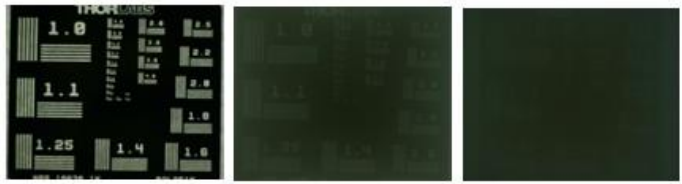

Fig. 11: Cell with dichroic molecules in the target setup at different voltages (0, 8 and 12 volts).

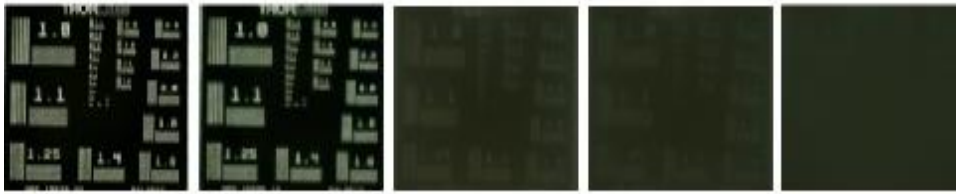

Fig. 12: Cell without dichroic molecules in the target setup at different voltages (0, 8, 12, 16 and 20 volts).

A salient feature of GH DSM mode compared to conventional PDLC/NCAP is that in case of power failure, the system returns to the clear state. This referred to as a "fail safe" performance is critical for many applications.

Table 1 shows a summary of characteristics and tradeoff between the different scattering systems. The optical clarity (low haze) in the off state, clear state transmission of >60%, low voltage requirements for complete opacity, and fail safe mode suggest that this system would be ideally suited for near eye applications. An example of this is eyewear for children's amblyopia where the visual acuity of the stronger eye is reduced to exercise the muscles of the weaker eye. In addition, this technology can be used for virtual reality applications.

Table 1: Comparison of different scattering systems.

|  | Fail Safe | Clear state haze | Drive voltage (V) | Turn off time | Off angle protection |
|---|---|---|---|---|---|
| PDLC | No | 15% | 50 | ms | No |
| Dye PDLC | No | 15% | 50 | ms | Yes |
| Normal mode PSCT | No | 7% | 40 | ms | No |
| Dye Normal mode PSCT | No | 7% | 40 | ms | Yes |
| Reverse mode PSCT | Yes | 7% | 30 | 0.5 s | No |
| Dye doped Reverse mode PSCT | No | 7% | 30 | 0.5 s | Yes |
| DSM | Yes | 1% | 12 | 1 s | No |
| **GH-DSM** | **Yes** | **1%** | **12** | **1 s** | **Yes** |

## 4. Conclusion:

We have constructed and characterized dye doped dynamic scattering system for controlling light transmission and scattering. Such devices will be useful for circumstances in which reducing the transmission of light and/or reducing visual acuity are required. The addition of dichroic dyes to dynamic scattering mode materials is necessary to give added control for both light absorption and scattering. Our statistical analysis shows that the optical properties of these devices can be tailored to have time-dependent attributes that are undetectable by human vision. For conditions requiring vision degradation, like amblyopia, the controllability of this device offers numerous advantages over traditional vision removal therapies such as eye patches.

**Acknowledgements**

We would like to thanks Dr. Ludmila Sukhomlinova for her fruitful suggestions. Also the Author feels particularly grateful to D. Gavazzi for his useful comments.